\documentclass[pra,superscriptaddress,twocolumn]{revtex4}
\usepackage{amssymb,amsthm,amsmath,color,graphics,epsfig}
\theoremstyle{plain}
\newtheorem{theorem}{Theorem}

\newtheorem{fact}[theorem]{Fact}

\theoremstyle{definition}
\newtheorem{definition}{Definition}

\newcommand{\ket}[1]{\ensuremath{\left|#1\right>}}
\newcommand{\bra}[1]{\ensuremath{\left<#1\right|}}
\newcommand{\braket}[2]{\ensuremath{\left<#1|#2\right>}}

\newcommand{\ketbra}[2]{| #1 \rangle\langle #2 |}

\newcommand{\be}{\begin{equation}}
\newcommand{\ee}{\end{equation}}
\newcommand{\bea}{\begin{eqnarray}}
\newcommand{\eea}{\end{eqnarray}}

\newcommand{\tr}{\textrm{tr}}
\renewcommand{\sp}{\textrm{span}}
\newcommand{\poly}{\text{poly}}

\def\opone{\leavevmode\hbox{\small1\kern-3.8pt\normalsize1}}

\newcommand{\dima}{M}
\newcommand{\dimb}{N}

\newcommand{\hermops}{\mathbb{H}_{\dima, \dimb}}

\newcommand{\densops}{\mathcal{D}_{\dima,\dimb}}

\newcommand{\sep}{\mathcal{S}_{\dima, \dimb}}
\newcommand{\ent}{\mathcal{E}_{\dima, \dimb}}

\newcommand{\n}{\dima^2\dimb^2}

\newcommand{\OSOPTK}{\mathcal{O}_{\text{SOPT}(K)}}

\newcommand{\OSSEPKprime}{\mathcal{O}_{\text{SSEP}(K')}}
\newcommand{\OSSEPKstar}{\mathcal{O}_{\text{SSEP}(K^\star)}}
\newcommand{\OSSEPQp}{\mathcal{O}_{\text{SSEP}(Q_p)}}

\begin{document}
\title{Quantum Separability and Entanglement Detection via Entanglement-Witness Search and Global Optimization}

\author{Lawrence M. Ioannou}
\affiliation{Centre for Quantum
Computation, Department of Applied Mathematics
  and Theoretical Physics, University of Cambridge, Wilberforce Road, Cambridge
  CB3 0WA, UK}

\author{Benjamin C. Travaglione}
\affiliation{Centre for Quantum
Computation, Department of Applied Mathematics
  and Theoretical Physics, University of Cambridge, Wilberforce Road, Cambridge
  CB3 0WA, UK}

\begin{abstract}
We focus on determining the separability of an unknown bipartite
quantum state $\rho$ by invoking a sufficiently large subset of
all possible entanglement witnesses given the expected value of
each element of a set of mutually orthogonal observables. We
review the concept of an entanglement witness from the geometrical
point of view and use this geometry to show that the set of
separable states is not a polytope and to characterize the class
of entanglement witnesses (observables) that detect entangled
states on opposite sides of the set of separable states. All this
serves to motivate a classical algorithm which, given the expected
values of a subset of an orthogonal basis of observables of an
otherwise unknown quantum state, searches for an entanglement
witness in the span of the subset of observables. The idea of such
an algorithm, which is an efficient reduction of the quantum
separability problem to a global optimization problem, was
introduced in PRA 70 060303(R), where it was shown to be an
improvement on the naive approach for the quantum separability
problem (exhaustive search for a decomposition of the given state
into a convex combination of separable states).  The last section
of the paper discusses in more generality such algorithms, which,
in our case, assume a subroutine that computes the global maximum
of a real function of several variables. Despite this, we
anticipate that such algorithms will perform sufficiently well on
small instances that they will render a feasible test for
separability in some cases of interest (e.g. in 3-by-3 dimensional
systems).
\end{abstract}

\maketitle

\section{Introduction}

\noindent Deciding whether a quantum state, be it physical or
theoretical, is separable (as opposed to entangled) is a problem
of fundamental importance in the field of quantum information
processing and is a computationally intractable problem
\cite{Gur03}.

One way to decide that a state is entangled is to use an
entanglement witness (EW) \cite{HHH96, Ter00}. Much work has been
done on entanglement witnesses (EWs) and their utility in
investigating the separability of quantum states, e.g.
\cite{LKCH00,LKHC01}. EWs have been found to be particularly
useful for experimentally detecting the entanglement of states of
the particular form $p\ketbra{\psi}{\psi}+(1-p)\sigma$, where
$\ket{\psi}$ is an entangled state and $\sigma$ is a mixed state
close to the maximally mixed state and $0\leq p\leq 1$
\cite{GHBELMS02,qphBMNMDM03}.

We will show that the set of separable states is not a polytope
and thus there is no finite set of EWs that can detect every
entangled state. This work focusses on the principle of invoking a
sufficiently large subset of all possible EWs given the expected
values of a set of observables. Section \ref{sec_Geo} summarizes
some geometric aspects of the set of Hermitian operators and
section \ref{sec_EWs} reviews the geometry of separable states and
entanglement witnesses. The simplest case of a set of expected
values giving rise to more than one EW is characterized in section
\ref{sec_AEW} and illustrated by visiting the problem of deciding
whether a noisy Bell state is entangled. In section
\ref{sec_DetectEntUnknown}, we apply the above principle to the
problem of detecting the entanglement of a completely unknown
quantum state and, in section \ref{sec_Alg}, we outline a class of
(classical) algorithms that search for an EW that detects the
state or conclude that so such EW exists (given the currently
available information about the state).

\section{Geometry of Vector Space of Hermitian Operators}\label{sec_Geo}

\noindent Let $\mathcal{H}_{M,N}$ denote the set of all Hermitian
operators mapping $\mathbb{C}^M\otimes\mathbb{C}^N$ to itself.
This vector space is endowed with the Hilbert-Schmidt inner
product $\langle X, Y \rangle\equiv \tr(AB)$, which induces the
corresponding norm $||X||\equiv\sqrt{\tr(X^2)}$ and distance
measure $||X-Y||$. By fixing an orthogonal Hermitian basis for
$\mathcal{H}_{M,N}$, the elements of $\mathcal{H}_{M,N}$ are in
one-to-one correspondence with the elements of the real Euclidean
space $\mathbb{R}^{M^2N^2}$.  Let
$\mathcal{B}=\{X_i:i=0,1,\ldots,\n-1\}$ be an orthonormal,
Hermitian basis for $\mathbb{H}_{M,N}$, where
$X_0\equiv\frac{1}{\sqrt{MN}}I$. For concreteness, we can assume
that the elements of $\mathcal{B}$ are tensor-products of the
(suitably normalized) canonical generators of SU(M) and SU(N),
given e.g. in \cite{TNWM02}. Note $\tr(X_i)=0$ for all $i>0$.
Define $v:\hermops\rightarrow \mathbb{R}^{\dima^2\dimb^2-1}$ as
\begin{eqnarray}\label{eqn_MappingFromHermopsToRealVecs}
v(A):=\begin{bmatrix} \tr(X_1 A) \\ \tr(X_2 A) \\
\vdots \\ \tr(X_{\dima^2\dimb^2-1} A)\end{bmatrix}.
\end{eqnarray}
Via the mapping $v$, the set of separable states $\sep$ can be
viewed as a full-dimensional convex subset of
$\mathbb{R}^{\dima^2\dimb^2-1}$
\begin{eqnarray}
\{v(\sigma)\in \mathbb{R}^{\dima^2\dimb^2-1}: \sigma\in\sep\},
\end{eqnarray}
which properly contains the origin
$v(I_{\dima,\dimb})=\overline{0}\in\mathbb{R}^{\dima^2\dimb^2-1}$
(recall that there is a ball of separable states of nonzero radius
centred at the maximally mixed state $I_{\dima,\dimb}$
\cite{GB02}).

Most of the definitions in the rest of this section may be found
in \cite{NW88}.  If $A\in \mathcal{H}_{M,N}$ and $A\neq 0$ and
$a\in\mathbb{R}$, then $\{x\in \mathcal{H}_{M,N}:\hspace{2mm}
\tr(Ax)\leq a\}$ is called the \emph{halfspace} $H_{A,a}$.  The
boundary $\{x\in \mathcal{H}_{M,N}:\hspace{2mm} \tr(Ax)=a\}$ of
$H_{A,a}$ is the \emph{hyperplane} $\pi_{A,a}$ with \emph{normal}
$A$.  Call two hyperplanes \emph{parallel} if they share the same
normal. Let $H^\circ_{A,a}$ denote the interior $H_{A,a}\setminus
\pi_{A,a}$ of $H_{A,a}$.  Note that $H^\circ_{-A,-a}$ is just the
complement of $H_{A,a}$. For example, the density operators of an
$M$ by $N$ quantum system lie on the hyperplane $\pi_{I,1}$, where
$I$ is the identity operator.  Let $\mathcal{D}_{M,N}=\{\rho\in
\mathcal{H}_{M,N}:\hspace{2mm} \rho\geq 0\}\cap \pi_{I,1}$ denote
the density operators.

The intersection of finitely many halfspaces is called a
\emph{polyhedron}.  Every polyhedron is a convex set.
 Let $D$ be a polyhedron. A set $F\subseteq D$ is a \emph{face} of
$D$ if there exists a halfspace $H_{A,a}$ containing $D$ such that
$F=D\cap \pi_{A,a}$. If $v$ is a point in $D$ such that the set
$\{v\}$ is a face of $D$, then $v$ is a \emph{vertex} of $D$.  A
\emph{facet} of $D$ is a nonempty face of $D$ having dimension one
less than the dimension of $D$.  A polyhedron that is contained in
a hyperball $\{x\in \mathcal{H}_{M,N}:\hspace{2mm} \tr(x^2)=
R^2\}$ of finite radius $R$ is a \emph{polytope}.

\section{Separable States and Entanglement Witnesses}\label{sec_EWs}
\noindent The set of bipartite separable quantum states $\sep$ in
$\mathcal{H}_{M,N}$ is defined as the convex hull of the separable
pure states
$\{\ketbra{\alpha}{\alpha}\otimes\ketbra{\beta}{\beta}\in
\mathcal{H}_{M,N}\}_{\alpha,\beta}$, where $\ket{\alpha}$ is a
unit vector in $\mathbb{C}^M$ and $\ket{\beta}$ is a unit vector
in $\mathbb{C}^N$. Let $\ent=\mathcal{D}_{M,N}\setminus \sep$ be
the set of entangled states.  For each entangled state $\rho$
there exists a halfspace $H_{A,a}$ whose interior $H^\circ_{A,a}$
contains $\rho$ but contains no member of $\sep$ \cite{HHH96}.
Call $A\in \mathcal{H}_{M,N}$ an \emph{entanglement witness}
\cite{Ter00} if for some $a\in\mathbb{R}$
\begin{eqnarray}\label{Def_LeftEW}
\sep\cap
H^\circ_{A,a}=\varnothing\hspace{3mm}\text{and}\hspace{3mm}\ent\cap
H^\circ_{A,a}\neq\varnothing.
\end{eqnarray}
Entanglement witnesses $A$ with $a=0$ in (\ref{Def_LeftEW})
correspond to the conventional definition of ``entanglement
witness'' found in the literature, e.g. \cite{GHBELMS02}.

Entanglement witnesses can be used to determine that a physical
quantum state is entangled. Suppose $A$ is an EW as in
(\ref{Def_LeftEW}) and that a state $\rho$ that is produced in the
lab is not known to be separable. If sufficiently many copies of
$\rho$ may be produced, then repeatedly measuring the observable
$A$ of $\rho$ gives a good estimate of the \emph{expected value of
A}
\begin{eqnarray}\nonumber
\langle A\rangle_\rho :=\tr(A\rho)
\end{eqnarray}
which, if less than $a$, indicates that $\rho\in H^\circ_{A,a}$
and hence that $\rho$ is entangled. Otherwise, if $\langle
A\rangle_\rho\geq a$, then $\rho$ may be entangled or separable.
The best value of $a$ to use in (\ref{Def_LeftEW}) is $a^* =
\min_{\ketbra{\psi}{\psi}\in \sep} \{\bra{\psi}A\ket{\psi}\}$
since, with this value of $a$, the hyperplane $\pi_{A,a}$ is
tangent to $\sep$ and thus the volume of entangled states that can
be detected by measuring observable $A$ is maximized.  With this
in mind, define
\begin{eqnarray}\nonumber
a^*(A):= \min_{\ketbra{\psi}{\psi}\in \sep}
\{\bra{\psi}A\ket{\psi}\}
\end{eqnarray}
if $A$ is an EW.

Detection of the entanglement of reproducible physical states in
the lab would be straightforward if there were a relatively small
number $K$ of EWs $A_i$ such that $\ent$ is contained in
\begin{eqnarray}\label{ENTComplementPolytope}\nonumber
\bigcup_{i=1}^{K} H_{A_i,a_i},
\end{eqnarray}
where $a_i:=a^*(A_i)$. This would imply that $\sep$ is
\begin{eqnarray}\label{SEPAPolytope}\nonumber
\bigcap_{i=1}^{K} H_{-A_i,-a_i},
\end{eqnarray}
that is, that $\sep$ is the intersection of finitely many
halfspaces.  Invoking the isomorphism between $\mathcal{H}_{M,N}$
and $\mathbb{R}^{M^2N^2}$, this says that $\sep$ is a polytope in
$\mathbb{R}^{M^2N^2-1}$.  Minkowski's theorem \cite{NW88} says
that every polytope in $\mathbb{R}^{n}$ is the convex hull of its
\emph{finitely many} vertices (extreme points).  Recall that an
extreme point of a convex set is one that cannot be written as a
nontrivial convex combination of other elements of the set.  To
show that $\sep$ is not a polytope, it suffices to show that it
has infinitely many extreme points.  The extreme points of $\sep$
are precisely the product states, as we now remind ourselves (see
also \cite{Hor97}):  A mixed state is not extreme, by definition.
Conversely, we have that
\begin{eqnarray}
\ketbra{\psi}{\psi}=\sum_ip_i\ketbra{\psi_i}{\psi_i}
\end{eqnarray}
implies
\begin{eqnarray}
1=\sum_ip_i\bra{\psi}\ketbra{\psi_i}{\psi_i}\ket{\psi}=\sum_ip_i
|\braket{\psi_i}{\psi}|^2,
\end{eqnarray}
which implies that $|\braket{\psi_i}{\psi}|=1$ for all $i$; thus,
a pure state is extreme.  Since $\sep$ has infinitely many pure
product states, we have the following fact, which settles a
problem mentioned in \cite{Bru02}.
\begin{fact}
$\sep$ is not a polytope in $\mathbb{R}^{\n-1}$.
\end{fact}

\section{Ambidextrous Entanglement Witnesses}\label{sec_AEW}

\noindent Suppose that $A$ is not an entanglement witness but that
$-A$ is. In this case, an estimate of $\tr(A\rho)$ is just as
useful in testing whether $\rho$ is entangled.  We extend the
definition of ``entanglement witness'' to reflect this fact:  Call
$A\in \hermops$ a \emph{left (entanglement) witness} if
(\ref{Def_LeftEW}) holds for some $a\in\mathbb{R}$, and a
\emph{right (entanglement) witness} if
\begin{eqnarray}\label{Def_RightEW}
\sep\cap
H^\circ_{-A,-b}=\varnothing\hspace{3mm}\text{and}\hspace{3mm}
\ent\cap H^\circ_{-A,-b}\neq\varnothing
\end{eqnarray}
for some $b\in\mathbb{R}$.  As well, for $A$ a right witness,
define
\begin{eqnarray}\nonumber
b^*(A):= \max_{\ketbra{\psi}{\psi}\in \sep}
\{\bra{\psi}A\ket{\psi}\}.
\end{eqnarray}
Note that $A$ is a left witness if and only if $-A$ is a right
witness.

The operator $A\in \hermops$ defines the family
$\{\pi_{A,a}\}_{a\in\mathbb{R}}$ of parallel hyperplanes in
$\mathbb{R}^{\n}$.  Consider the hyperplane
$\pi_A:=\pi_{A,\frac{\tr(A)}{MN}}$ which cuts through $\sep$ at
the maximally mixed state $I_{MN}$.  When can $\pi_A$ be shifted
parallel to its normal so that it separates $\sep$ from some
entangled states? If $A$ is \emph{both} a left and right witness,
then $\pi_A$ can be shifted either in the positive or negative
directions of the normal.  In this case, the two parallel
hyperplanes $\pi_{A,a^*(A)}$ and $\pi_{A,b^*(A)}$ sandwich $\sep$
with some entangled states outside of the \emph{sandwich}, which
we will denote by $W(A):=H_{-A,-a^*(A)}\cap H_{-A,-b^*(A)}$.
\begin{definition}[Ambidextrous entanglement witness]
An operator $A\in \hermops$ is an \emph{ambidextrous
(entanglement) witness} if it is both a left witness and a right
witness.
\end{definition}
If $A$ is an ambidextrous witness, then $\rho$ is entangled if
$\langle A\rangle_\rho <a^*(A)$ \emph{or} if $\langle
A\rangle_\rho
>b^*(A)$.  We can further define a \emph{left-handed} witness to
be an entanglement witness that is left but not right. Say that
two entangled states $\rho_1$ and $\rho_2$ are \emph{on opposite
sides of $\sep$} if there does not exist a halfspace $H_{A,a}$
such that $H^\circ_{A,a}$ contains $\rho_1$ and $\rho_2$ but
contains no separable states.  Ambidextrous witnesses have the
potential advantage over conventional (left-handed) entanglement
witnesses that they can detect entangled states on opposite sides
of $\sep$ with the \emph{same} physical measurement.

Entanglement witnesses can be simply characterized by their
spectral decomposition. In the following, suppose $A\in \hermops$
has spectral decomposition $A=\sum_{i=0}^{MN-1} \lambda_i
\ketbra{\lambda_i}{\lambda_i}$ with
$\lambda_0\leq\lambda_1\leq\ldots\leq\lambda_{MN-1}$.
\begin{fact}\label{Fact_CharShiftLeft}
The operator $A$ is a left witness if and only if there
 exists $k\in [0,1,\ldots,MN-2]$ such that
 $\sp(\{\ket{\lambda_0},\ket{\lambda_1},\ldots,\ket{\lambda_k}\})$
 contains no separable pure states and $\lambda_{k+1}>\lambda_k$.
\end{fact}

\begin{proof}Suppose first that there exists no such $k$. Then
$\ket{\lambda_0}$ is, without loss of generality, a separable pure
state (because the eigenspace corresponding to $\lambda_0$ must
contain a product state), so $A$ cannot be a left witness. To
prove the converse, suppose that such a $k$ does exist and that
$\lambda_{k+1}>\lambda_k$. Define the real function
$f(\sigma):=\tr(A\sigma)$ on $\sep$. Since
$\sp(\{\ket{\lambda_0},\ket{\lambda_1},\ldots,\ket{\lambda_k}\})$
contains no separable states and $\lambda_{k+1}>\lambda_k$, the
function satisfies $f(\sigma)>\lambda_0$. Since the set of
separable states is compact, there exists a separable state
$\sigma'$ that minimizes $f(\sigma)$. Thus, setting
$a:=f(\sigma')$ gives $\sep\cap H^\circ_{A,a}=\varnothing$. As
well, $\ent\cap H^\circ_{A,a}\neq\varnothing$ since
$\tr(A\ketbra{\lambda_0}{\lambda_0})=\lambda_0<a$, and so $A$ is a
left witness.
\end{proof}

\begin{theorem}\label{Theorem_CharShift}
The operator
 $A$ is a left or right entanglement witness if and only if (i) there
 exists $k\in [0,1,\ldots,MN-2]$ such that
 $\sp\{\ket{\lambda_0},\ket{\lambda_1},\ldots,\ket{\lambda_k}\}$
 contains no separable pure states and $\lambda_{k+1}>\lambda_k$, or (ii) there
 exists $l\in [1,2,\ldots,MN-1]$ such that\\
 $\sp\{\ket{\lambda_l},\ket{\lambda_{l+1}},\ldots,\ket{\lambda_{MN-1}}\}$
 contains no separable pure states and $\lambda_{l}>\lambda_{l-1}$.
\end{theorem}

Theorem \ref{Theorem_CharShift} immediately gives a method for
identifying and constructing entanglement witnesses.
\begin{definition}[Partial Product Basis, Unextendible Product Basis \cite{Ter01}] A \emph{partial product
basis} of $\mathbb{C}^\dima\otimes\mathbb{C}^\dimb$ is a set $S$
of mutually orthonormal pure product states spanning a proper
subspace of $\mathbb{C}^\dima\otimes\mathbb{C}^\dimb$.  An
\emph{unextendible product basis} of
$\mathbb{C}^\dima\otimes\mathbb{C}^\dimb$ is a partial product
basis $S$ of $\mathbb{C}^\dima\otimes\mathbb{C}^\dimb$ whose
complementary subspace $(\sp S)^\perp$ contains no product state.
\end{definition}
\noindent We can use unextendible product bases to construct
ambidextrous witnesses.  Suppose $B$ is an unextendible product
basis of $\mathbb{C}^\dima\otimes\mathbb{C}^\dimb$, and let $B'$
be disjoint from $B$ such that $B\cup B'$ is an orthonormal basis
of $\mathbb{C}^\dima\otimes\mathbb{C}^\dimb$. One possibility is
the left witness defined by $A'$ as
\begin{eqnarray}
A' = -\sum_{\ket{\lambda}\in B'}\ketbra{\lambda}{\lambda}
\end{eqnarray}
As well, we could split $B'$ into $B'_L$ and $B'_R$ and define an
ambidextrous witness $A''$ as
\begin{eqnarray}
A'' = -\sum_{\ket{\lambda_L}\in B'_L}\ketbra{\lambda_L}{\lambda_L}
+\sum_{\ket{\lambda_R}\in B'_R}\ketbra{\lambda_R}{\lambda_R}.
\end{eqnarray}
\noindent Another thing to realize is that $\sp B$ may contain an
entangled pure state, which can be pulled out and put into a
$(+1)$-eigenvalue eigenspace of $A'$. Depending on $B$ (and the
dimensions $\dima$, $\dimb$), there may be several mutually
orthogonal pure entangled states in $\sp B$ whose span contains no
product state; let $B''$ be a set of such pure states.  Define the
ambidextrous witness as
\begin{eqnarray}
A''' = -\sum_{\ket{\lambda}\in B'}\ketbra{\lambda}{\lambda} +
\sum_{\ket{\lambda}\in B''}\ketbra{\lambda}{\lambda}.
\end{eqnarray}
\noindent This suggests the following problem, related to the
combinatorial \cite{AL01} problem of finding unextendible product
bases: Given $\dima$ and $\dimb$, find all orthonormal bases $B$
for $\mathbb{C}^\dima\otimes\mathbb{C}^\dimb$ such that
\begin{itemize}
\item $B$ is the disjoint union of $\Lambda_L$, $\tilde{B}$,
$\Lambda_R$, \item $\sp\Lambda_L$ and $\sp\Lambda_R$ contain no
product state, \item $\sp( \Lambda_L\cup\Lambda_R)$ contains a
product state, and \item $\min\{|\Lambda_L|,|\Lambda_R|\}$ is
maximal.
\end{itemize}
\noindent Such bases may give ``optimal'' ambidextrous witnesses,
which detect the largest volume of entangled states on opposite
sides of $\sep$.

The functions $a^*$ and $b^*$ are difficult to compute
\footnote{At least, (see Section \ref{sec_Alg}) WOPT($\sep$) is an
NP-hard problem, because WSEP($\sep$) is both NP-hard \cite{Gur03,
GLS88} and, as the existence of the algorithms described in
Section \ref{sec_Alg} proves, efficiently reducible to
WOPT($\sep$).}. Thus a criticism of constructing witnesses via the
spectral decomposition is that even if you can construct the
corresponding physical observables, you still have to perform a
difficult computation to make them useful. However, most
experimental applications of entanglement witnesses are in very
low dimensions, where computing $a^*$ and $b^*$ deterministically
is not a problem -- it may even be done analytically, as in the
example at the end of this section.

Ambidextrous witnesses represent the simplest case of the
principle of invoking as many (left) entanglement witnesses as
possible given the expected values of each element of a set $X$ of
linearly independent observables, that is, the case $|X|=1$.  In
the Section \ref{sec_DetectEntUnknown}, we see how this principle
generalizes to $|X|>1$.

A simple illustration of how AEWs may be used involves detecting
and distinguishing noisy Bell states. Define the four Bell states
in $\mathbb{C}^2\otimes\mathbb{C}^2$:
\begin{eqnarray}\nonumber
\ket{\psi^\pm}&:=&\left(\ket{00}\pm\ket{11}\right)/\sqrt{2}\\\nonumber
\ket{\phi^\pm}&:=&\left(\ket{01}\pm\ket{10}\right)/\sqrt{2}.
\end{eqnarray}

It is straightforward to show that the Bell states are, pairwise,
on opposite sides of $\mathcal{S}_{2,2}$.  Suppose a left
entanglement witness $W$, with $a^*(W)=0$, detects $\ket{\psi^+}$
and $\ket{\phi^+}$.  Without loss of generality, $W$ can be
written in the Bell basis $\{  \ket{\psi^+}, \ket{\phi^+}, \ldots
\}$ as
\begin{eqnarray}W=\begin{bmatrix}-\epsilon_1 & a+bi & \times & \times \\
a-bi & -\epsilon_2 & \times & \times
\\ \times & \times & \times & \times \\ \times & \times & \times & \times \\ \end{bmatrix},\end{eqnarray}
for $\epsilon_1$ and $\epsilon_2$ both positive.  But the states
$\ket{s^\pm}\equiv\frac{1}{\sqrt{2}}(\ket{\psi^+}\pm\ket{\phi^+})$
are separable.  Requiring $\bra{s^+}W\ket{s^+}\geq 0$ gives
$2a\geq\epsilon_1+\epsilon_2$ and requiring
$\bra{s^-}W\ket{s^-}\geq 0$ gives $2a\leq -\epsilon_1-\epsilon_2$,
which, together, give a contradiction.  Similar arguments hold for
the other pairs of Bell states.

Define the operators
\begin{eqnarray}\nonumber
A_\psi&:=& -\ketbra{\psi^-}{\psi^-} +
\ketbra{\psi^+}{\psi^+}\\\nonumber A_\phi&:=&
-\ketbra{\phi^-}{\phi^-} + \ketbra{\phi^+}{\phi^+}.
\end{eqnarray}
Both $A_\psi$ and $A_\phi$ are easily seen to be AEWs.  It is also
straightforward to compute the values
\begin{eqnarray}\nonumber
a^*(A_\psi)=a^*(A_\phi)= -1/2
\end{eqnarray}
and
\begin{eqnarray}\nonumber
 b^*(A_\psi)=b^*(A_\phi)=
+1/2.
\end{eqnarray}
Suppose that there is a source that repeatedly emits the same
noisy Bell state $\rho$ and that we want to decide whether $\rho$
is entangled.  Define the Pauli operators:
\begin{equation}\label{eqn_PauliOperators}
\begin{array}{ccrccrl}
\nonumber \sigma_0 &:=&   \frac{1}{\sqrt{2}} (\ketbra{0}{0} +
\ketbra{1}{1})&\\
\nonumber \sigma_1 &:=&   \frac{1}{\sqrt{2}}(\ketbra{0}{1}+ \ketbra{1}{0})&\\
\nonumber \sigma_2 &:=&- \frac{i}{\sqrt{2}}(\ketbra{0}{1}- \ketbra{1}{0})&\\
\nonumber \sigma_3 &:=&   \frac{1}{\sqrt{2}}(\ketbra{0}{0} -
\ketbra{1}{1})&,
\end{array}
\end{equation}
where $\{\ket{0},\ket{1}\}$ is the standard orthonormal basis for
$\mathbb{C}^2$.  Noting that
\begin{eqnarray}\nonumber
A_\psi&=& \sigma_1\otimes\sigma_1 -
\sigma_2\otimes\sigma_2\\
\nonumber A_\phi&=&\sigma_1\otimes\sigma_1 +
\sigma_2\otimes\sigma_2,
\end{eqnarray}
measuring the expected value of the two observables
$\sigma_1\otimes\sigma_1$ and $\sigma_2\otimes\sigma_2$ may be
sufficient to decide that $\rho$ is entangled because $\rho\in
\mathcal{E}_{2,2}$ if one of the following four inequalities is
true:
\begin{eqnarray}\label{ineq_SuffEnt}
\langle \sigma_1\otimes\sigma_1 \rangle_\rho\pm
\langle\sigma_2\otimes\sigma_2\rangle_\rho&>&1/2 \\\nonumber
\langle \sigma_1\otimes\sigma_1 \rangle_\rho\pm
\langle\sigma_2\otimes\sigma_2\rangle_\rho&<&-1/2 .
\end{eqnarray}
If the noise is known to be of a particular form, then we can also
determine \emph{which} noisy Bell state was being produced. Let
$\ket{B}$ be a Bell state. Suppose $\rho$ is known to be of the
form $p\ketbra{B}{B}+(1-p)\sigma$ for some $\sigma$ inside both
sandwiches $W(A_\psi)$ and $W(A_\phi)$. With $\sigma$ so defined,
one of the four inequalities (\ref{ineq_SuffEnt}) holds only if
exactly one of them holds, so that $\ket{B}$ is determined by
which inequality is satisfied. We remark that, if $\sigma$ and
$\ket{B}$ are known, knowledge of the expected value of any single
observable $A$ may allow one to compute $p$ and hence an upper
bound on the $l_2$ distance between $\rho$ and the maximally mixed
state $I/4$. This distance may be enough information to conclude
that $\rho$ is separable by checking if $\rho$ is inside the
largest separable ball centred at $I/4$ \cite{GB02}.

\section{Detecting Entanglement of an Unknown
State Using Partial Information}\label{sec_DetectEntUnknown}

\noindent We now consider the task of trying to decide whether a
completely unknown physical state $\rho$, of which many copies are
available, is entangled.  For simplicity, we restrict to
$\rho\in\mathcal{H}_{2,2}$ but the discussion can be applied to a
bipartite system of any dimension, replacing Pauli operators with
canonical generators of SU(M) and SU(N) (or any linearly
independent Hermitian product basis). For such $\rho$, this
problem has already been addressed in \cite{HE02}, where the
so-called ``structural physical approximation of an unphysical
map'' \cite{qphHor01} was used to implement the Peres-Horodecki
positive partial transpose (PPT) test \cite{Per96,HHH96}.  While
the structural physical approximation is experimentally viable in
principle, it is currently very difficult to do so. Thus, the
easiest way to test for entanglement at present is to perform
``state tomography'' in order to get good estimates of 15 real
parameters that define $\rho$, then reconstruct the density matrix
for $\rho$ and carry out the PPT test \cite{Per96} on this matrix.

An experimentalist has many choices of which 15 parameters to
estimate: the expectations of any 15 linearly independent
observables qualify, as do the probability distributions of any 5
mutually unbiased (four-outcome) measurements \cite{Iva81,WF89}.
Whatever 15 parameters are chosen, we assume that the basic tool
of the experimentalist is the ability to perform local two-outcome
measurements on each qubit, e.g. measuring $\sigma_1$ on the first
qubit and $\sigma_2$ on the second.  Under this assumption, the
scenario where the two qubits of $\rho$ are far apart is easily
handled if classical communication is allowed between the two
labs.  We further assume, for simplicity, that the set of these
local two-outcome measurements is the set of Pauli operators
$\{\sigma_i\}_{i=0,1,2,3}$ (defined on page
\pageref{eqn_PauliOperators}). If $\sigma_i$ is measured on the
first qubit and $\sigma_j$ on the second, repeating this procedure
on many copies of $\rho$ gives good estimations of the three
expectations $\langle\sigma_i\otimes\sigma_0\rangle$,
$\langle\sigma_0\otimes\sigma_j\rangle$, and
$\langle\sigma_i\otimes\sigma_j\rangle$ (where the subscript
``$\rho$'' is omitted for readability). Let us call this procedure
\emph{measuring $\sigma_i\sigma_j$}.

Suppose the experimentalist sets out to solve our problem and
begins the data collection by measuring $\sigma_1\sigma_1$ and
then $\sigma_2\sigma_2$.  Even though only 6 of the 15 independent
parameters defining $\rho$ have been found, the example in the
previous section shows that $\rho$ is entangled if one of the four
inequalities (\ref{ineq_SuffEnt}) is true. It is straightforward
to show that if none of these inequalities is true, then no
entanglement witness in the span of
$\{\sigma_1\otimes\sigma_1,\sigma_2\otimes\sigma_2\}$ can detect
$\rho$ if it is entangled \footnote{To show this, it suffices to
find four separable states whose projections onto
$\sp\{\sigma_1\otimes\sigma_1,\sigma_2\otimes\sigma_2\}$ are the
four vertices of the square with vertices $(\frac{1}{2},0)$,
$(0,\frac{1}{2})$, $(-\frac{1}{2},0)$, and $(0,-\frac{1}{2})$;
such states are $\frac{1}{4}I \pm
\frac{1}{2}\sigma_i\otimes\sigma_i$ for $i=1,2$. The result then
follows from convexity of $\mathcal{S}_{2,2}$.}. However, there
may be an entanglement witness in the span of
\begin{eqnarray}\nonumber
\{\sigma_0\otimes\sigma_1, \sigma_0\otimes\sigma_2,
\sigma_1\otimes\sigma_1, \sigma_2\otimes\sigma_2,
\sigma_1\otimes\sigma_0, \sigma_1\otimes\sigma_0\}
\end{eqnarray}
that does detect $\rho$.

More generally, at any stage of the data-gathering process, if we
have the set of expectations
$\{\langle\sigma_i\otimes\sigma_j\rangle: (i,j)\in T\}$, then
$\rho$ is entangled if there is an entanglement witness in the
span of $\{\sigma_i\otimes\sigma_j: (i,j)\in T\}$ that detects
$\rho$ ($T\subset \{(k,l):k,l\in\{0,1,2,3\}\}\setminus (0,0)$). If
the experimentalist has access to a computer program that can
quickly discover such an entanglement witness (if it exists), then
the data-gathering process can be terminated early and no more
qubits have to be used to decide that $\rho$ is entangled. The
algorithms described in the next section are just such programs.
To see this, note that the projection
$\overline{\mathcal{S}}_{2,2}$ of $\mathcal{S}_{2,2}$ onto
$\sp\{\sigma_i\otimes\sigma_j: (i,j)\in T\}$ is a full-dimensional
convex subset of $\mathbb{R}^{|T|}$, and the projection
$\overline{\rho}$ of $\rho$ onto $\sp\{\sigma_i\otimes\sigma_j:
(i,j)\in T\}$ is a point in $\mathbb{R}^{|T|}$ such that
$\overline{\rho}\notin\overline{\mathcal{S}}_{2,2}$ if and only if
there is an entanglement witness in the span of
$\{\sigma_i\otimes\sigma_j: (i,j)\in T\}$ that detects $\rho$.
Since the following algorithms can be applied to any
full-dimensional convex set (satisfying certain conditions), we
can apply them to $\overline{\mathcal{S}}_{2,2}$.

We view any such algorithm as an extra tool that an
experimentalist can use to facilitate entanglement detection and
minimize the number of copies of $\rho$ that must be measured --
essentially, trading classical resources for quantum resources.
As we saw in the case of constructing ambidextrous witnesses, the
primary classical resource required to invoke a sufficiently large
subset of all such entanglement witnesses is a subroutine for
computing the function $b^*$ (equivalently, $a^*$).

\section{Algorithms for Finding Entanglement Witnesses Based on Global Optimization}\label{sec_Alg}

\noindent Assume that $\rho\in\mathcal{D}_{M,N}$ is a state whose
separability is unknown.  We can handle two scenarios -- one
experimental, as described above, and the other theoretical.  In
the theoretical scenarios, we assume that we know the density
matrix for $\rho$; this corresponds to having gathered all
$M^2N^2-1$ independent expected values in the experimental
scenario.  Since the algorithms find an entanglement witness when
$\rho\in \ent$, they could also be applied when $\rho$ is known to
be entangled but an entanglement witness for $\rho$ is desired
(though one may want to apply the entanglement witness
optimization procedure \cite{LKCH00} to the result of the
algorithm, as these algorithms do not necessarily output optimal
entanglement witnesses).

Let $j$ be the number of nontrivial expected values of $\rho$ that
are known, $2\leq j\leq M^2N^2-1$; that is, (without loss) assume
we know the expected values of the elements of
$\mathcal{B}'=\{X_1, X_2,\ldots, X_j\}$.  The algorithms either
find an entanglement witness in $\sp(\mathcal{B}')$ for $\rho$, or
conclude that no such witness exists.  For any
$Y\in\mathbb{H}_{M,N}$ with $Y=\sum_{i=0}^{M^2N^2-1} y_iX_i$, let
$\overline{Y}$ be the $j$-dimensional vector of the real numbers
$y_i$ for $i=1,2,\ldots,j$. Define
\begin{eqnarray}
\overline{\mathcal{S}}_{M,N} = \{\overline{\sigma}: \sigma\in
\sep\}.
\end{eqnarray}
\noindent Note $\overline{\mathcal{S}}_{M,N}$ is a
full-dimensional convex set in $\mathbb{R}^j$, properly containing
the origin (since $\overline{I_{M,N}}$ is the zero-vector in
$\mathbb{R}_j$).

Let $K$ be a full-dimensional convex subset of $\mathbb{R}^n$
which contains a ball of finite nonzero radius centred at the
origin and is contained in a ball of finite radius $R$.  The
algorithms are general and can be used to decide whether a
hyperplane exists which separates a given point $p$ from any given
$K$ satisfying these properties.  Thus, the clearest way to
describe how the algorithms work is to use the application-neutral
notation of convex analysis.  For $x\in\mathbb{R}^n$ and
$\delta>0$, let $B(x,\delta):= \{y\in\mathbb{R}^n:
||x-y||\leq\delta\}$.  For a convex subset $K\subset\mathbb{R}^n$,
let $S(K,\delta):=\cup_{x\in K} B(x,\delta)$ and
$S(K,-\delta):=\{x: B(x,\delta)\subseteq K\}$.
 Define the following convex body problems \cite{GLS88}:

\begin{definition}[Weak separation problem for $K$ ($\text{WSEP($K$)}$)]\label{def_WSEP}
Given a rational vector $p\in\mathbb{R}^n$ and rational
$\delta>0$, either
\begin{itemize}
\item assert $p\in S(K,\delta)$,
\hspace{2mm}\text{or}\label{eqn_WSEPSepAssertion}

\item find a rational vector $c\in\mathbb{R}^n$ with
$||c||_\infty= 1$ such that $c^Tx < c^Tp$ for every $x\in K$
\footnote{The $l_\infty$ norm appears here as a technicality, so
that $c$ need not be normalized by a possibly irrational
multiplier.  We will just use the Euclidean norm in what follows
and have $||c|| \approx 1$.}.\label{eqn_WSEPEntAssertion}
\end{itemize}
\end{definition}

\begin{definition}[Weak optimization problem for $K$ (WOPT($K$))]\label{def_WOPT}
Given a rational vector $c\in\mathbb{R}^n$ and rational
$\epsilon>0$, either
\begin{itemize}
\item find a rational vector $y\in\mathbb{R}^n$ such that $y\in
S(K,\epsilon)$ and  $c^Tx\leq c^Ty +\epsilon$ for every $x\in K$;
or

\item assert that $S(K,-\epsilon)$ is empty \footnote{This will
never be the case for us, as $\sep$ is not
empty.}.\label{eqn_WSEPEntAssertion}
\end{itemize}
\end{definition}

\noindent By taking $\delta$ and $\epsilon$ to be zero, we
implicitly define the corresponding \emph{strong} separation
(SSEP) and \emph{strong} optimization (SOPT) problems.  Note that
by taking $K$ to be $\overline{\mathcal{S}}_{M,N}$ and $p$ to be
$\overline{\rho}$ (for some state $\rho\in\densops$), SSEP($K$)
corresponds to the problem of finding an entanglement witness for
$\rho$ (or deciding that one does not exist in the span of
$\mathcal{B'}$) \footnote{If $p$ arises from some estimation
procedure (as in our experimental setting), then there is a
hyperbox around $p$ that contains the ``actual'' point
$p^\checkmark$; the hyperbox is given by the ``error bars'' on
each coordinate of $p$. From the ``error bars'' can be computed a
$\Delta>0$ such that $p^\checkmark\in B(p,\Delta)$.  If the
WSEP($K$) algorithm asserts $p\in S(K,\delta)$, then we can only
assert that $p^\checkmark$ is in $S(K,\delta+\Delta)$; otherwise,
we can only assert that $c^Tx < c^Tp^\checkmark +\Delta$ for every
$x\in K$.}; and by further taking $c$ to be $\overline{A}$, for
some $A\in\hermops$, SOPT($K$) corresponds to the problem of
computing $b^*(A)$ (actually something at least as hard, since
SOPT($K$) asks for the \emph{maximizer} of $c^Tx$, over $x\in K$,
rather than just the \emph{maximum}).

We describe \emph{oracle-polynomial-time} algorithms for
WSEP($K$), with respect to an oracle for WOPT($K$); that is,
assuming each call to the oracle is assigned unit complexity cost,
the algorithms have running time in $O(\poly(n,\log(R/\delta)))$.
We will use ``$\mathcal{O}$'' to denote oracles (black-boxed
subroutines) for problems, indicating which problem via a
subscript, e.g. $\OSOPTK$.  In what follows, so as not to
obfuscate the main idea of the algorithms, we ignore the weakness
of the separation and optimization problems; that is, we assume we
are solving SSEP($K$) with an oracle for SOPT($K$).

There are at least two ways to reduce SSEP($K$) to SOPT($K$).  The
first method was covered in \cite{ITCE04, qphIoa05}; the second
method, which we give below, is well known and may be found in the
synthesis of Lemma 4.4.2 and Theorem 4.2.2 in \cite{GLS88}. For
$y\in\mathbb{R}^n$ and $b\in\mathbb{R}$, define the hyperplane
$\pi_{y,b}\equiv \{x\in\mathbb{R}^n: y^Tx=b\}$.

\begin{definition}[Polar of $K$] The \emph{polar} $K^\star$ of a full-dimensional convex set
$K\subset \mathbb{R}^n$ that contains the origin is defined as
\begin{eqnarray}
K^\star := \{c\in\mathbb{R}^n: c^Tx\leq 1 \hspace{2mm}\forall x
\in K\}.
\end{eqnarray}
\end{definition} \noindent If $c\in K^\star$, then the plane $\pi_{c,1}$ separates $p\in\mathbb{R}^n$ from $K$ when $c^Tp >1$.
\begin{definition}[Feasibility problem for $K'$ (FEAS($K'$))]\label{def_FEAS}
Given a convex set $K'\subset\mathbf{R}^n$, either
\begin{itemize}
\item find a point $k'\in K'$, or

\item assert that $K'$ is empty.
\end{itemize}
\end{definition}
\noindent Thus, the separation problem for $p$ is equivalent to
the feasibility problem for $Q_p$, defined as
\begin{eqnarray}
Q_p := K^\star \cap \{c: p^Tc \geq 1\}.
\end{eqnarray}
As outlined in the next section, to solve the feasibility problem
for any $K'$, it suffices to have a separation routine for $K'$.
Because we can easily build a separation routine $\OSSEPQp$ for
$Q_p$ out of $\OSSEPKstar$, it suffices to have a separation
routine $\OSSEPKstar$ for $K^\star$ in order to solve the
feasibility problem for $Q_p$ \footnote{We slightly abuse the
oracular ``$\mathcal{O}$'' notation by using it for both truly
oracular (black-boxed) routines and for other (possibly not
completely black-boxed) routines.}.  Building $\OSSEPQp$ out of
$\OSSEPKstar$ is done as follows:\\

\noindent Routine $\OSSEPQp(y)$:\\
\noindent \textsc{case:} $p^Ty < 1$\\
\noindent \indent \textsc{return} $-p$\\
\noindent \textsc{else:} $p^Ty \geq 1$\\
\noindent \indent \textsc{call} $\OSSEPKstar(y)$\\
\noindent \indent \textsc{case:} $\OSSEPKstar(y)$ returns separating vector $q$\\
\noindent \indent\indent \textsc{return} $q$\\
\noindent \indent \textsc{else:} $\OSSEPKstar(y)$ asserts $y\in K^\star$\\
\noindent \indent\indent \textsc{return} ``$y\in Q$''\\

\noindent It remains to show that the optimization routine
$\OSOPTK$ for $K$ gives a separation routine $\OSSEPKstar$ for
$K^\star$. Suppose $y$ is given to $\OSOPTK$, which returns $k\in
K$ such that $y^Tx \leq y^Tk =:b$ for all $x\in K$. If $b\leq 1$,
then $\OSSEPKstar$ may assert $y\in K^\star$. Otherwise,
$\OSSEPKstar$ may return $k$, because $\pi_{k,1}$ (and hence
$\pi_{k,b}$) separates $y$ from $K^\star$: since $k^Ty=b > 1$, it
suffices to note that $k^Tc=c^Tk\leq 1$ for all $c\in K^\star$ by
the definition of $K^\star$ and the fact that $k\in K$.

The plane $\pi_{k,1}$ is called a \emph{cutting plane}, and, to
solve FEAS($K'$) with $\OSSEPKprime$, we use a \emph{cutting-plane
algorithm}. All such algorithms have the same basic structure:

\begin{enumerate}

\item Define a (possibly very large) regular bounded convex set
$P_0$ which is guaranteed to contain $K'$, such that, for some
reasonable definition of ``centre'', the centre $\omega_0$ of
$P_0$ is easily computed.  The set $P_0$ is called an \emph{outer
approximation} to $K'$.  Common choices for $P_0$ are the
origin-centred hyperbox, $\{x\in\mathbb{R}^n: -2^L \leq x_i \leq
2^L, \hspace{2mm}1\leq i\leq n \}$ and the origin-centred
hyperball, $\{x: x^Tx \leq 2^L\}$ (where $2^L$ is a trivially
large bound).

\item Give the centre $\omega$ of the current outer approximation
$P$ to $\OSSEPKprime$.

\item If $\OSSEPKprime$ asserts ``$\omega\in K'$'', then HALT.

\item Otherwise, say $\OSSEPKprime$ returns the cutting plane
$\pi_{c,b}$ such that $K'\subset \{x: c^Tx \leq b\}$. Update
(shrink) the outer approximation $P:=P\cap \{x: c^Tx \leq b'\}$
for some $b'\geq b$; the idea is that the new $P$ has about half
the volume of the old $P$ (i.e. usually $\pi_{c,b}$ passes through
$\omega$, or nearby it). Possibly perform other computations to
further update $P$. Check stopping conditions; if they are met,
then HALT. Otherwise, go to step 2.

\end{enumerate}
\noindent  The difficulty with such algorithms is knowing when to
halt in step 4.  Generally, the stopping conditions are related to
the size of the current outer approximation.  Because it is always
an approximate (weak) feasibility problem that is solved, the
associated accuracy parameter $\delta'$ can be exploited to get a
``lower bound'' $V$ on the ``size'' of $K'$, with the
understanding that if $K'$ is smaller than this bound, then the
algorithm can correctly assert that $S(K',-\delta')$ is empty.
Thus the algorithm stops in step 4 when the current outer
approximation is smaller than $V$.


Using the above reduction from SSEP($K$) to SOPT($K$), there are a
number of polynomial-time convex feasibility algorithms that can
be applied (see \cite{AV95} for a discussion of all of them). The
three most important are the \emph{ellipsoid method}, the
\emph{volumetric-centre method}, and the \emph{analytic-centre
method}.
The latter are more efficient than the ellipsoid algorithm and are
very similar to each other in complexity and precision
requirements, with the analytic-centre cutting-plane (ACCP)
algorithm in \cite{AV95} having some supposed practical
advantages.

We refer to \cite{qphIoa05} and \cite{AV95} (and references
therein) for a discussion of details, including computer precision
requirements, for either of the ACCP algorithms arising from
either the reduction in \cite{ITCE04} or the well-known one given
here.

\section{Closing Remarks}

\noindent In terms of attempting to find a practical algorithm for
the quantum separability problem, the skeptic notices that such
algorithms appear not to offer any advantage over other
approaches: instead of having to solve one instance of an NP-hard
problem, we now have to solve many!  In response to such
skepticism, we can, at least, refer to \cite{qphIoa06}, where it
is shown that the asymptotic complexity of such algorithms
compares well with all other deterministic algorithms (with known
worst-case complexity bounds) for the quantum separability
problem.  There are many algorithms available for optimizing
functions and thus computing WOPT($\sep$), including the
semidefinite programming relaxation method of Lasserre
\cite{Las01} (on which, incidentally, one can base a different
separability algorithm \cite{EHGC04}); Lipschitz optimization
\cite{HP95}; and Hansen's global optimization algorithm using
interval analysis \cite{Han92}.

\section{Acknowledgements}

\noindent We would like to thank Carolina Moura Alves, Coralia
Cartis, and Tom Stace for helpful discussions and assistance. This
work was supported by the CESG, NSERC, ORS, RESQ (EU grant
IST-2001-37559), and CMI.


\end{document}